\documentclass[onecolumn]{emulateapj}

\usepackage[bookmarks,breaklinks]{hyperref}
   \hypersetup{
     colorlinks,
     citecolor=blue,
     linkcolor=blue,
    }

\usepackage{graphicx,times}             %for PS/EPS graphics inclusion, new
\usepackage{rotating}

\shorttitle{Study on the energy limits of kHz QPOs in Sco X-1 with $RXTE$ and $Insight$-HXMT observations}
\shortauthors{Jia et al.}

\received{}
\revised{}
\accepted{}

\begin{document}

\title{Study on the energy limits of kHz QPOs in Sco X-1 with $RXTE$ and $Insight$-HXMT observations}

\author{
S. M. Jia$^{1,2}$, J. L. Qu$^{1,2}$, F. J. Lu$^{1,3}$, S. N. Zhang$^{1,2}$, S. Zhang$^{1}$,
Y. Huang$^{1}$, D. H. Wang$^{4}$, D. K. Zhou$^{1}$, G. C. Xiao$^{5}$,
Q.C. Bu$^{1,6}$, L. Chen$^{7}$, X. Ma$^{1}$, L. M. Song$^{1,2}$, L. Tao$^{1}$, X. L. Cao$^{1}$, Y. Chen$^{1}$, C. Z. Liu$^{1}$, Y. P. Xu$^{1,2}$
}

\altaffiltext{1}{Key Laboratory of Particle Astrophysics, Institute of High Energy Physics, Chinese Academy of Sciences, Beijing 100049, China}
\altaffiltext{2}{University of Chinese Academy of Sciences, Chinese Academy of Sciences, Beijing 100049, China}
\altaffiltext{3}{Key Laboratory of Stellar and Interstellar Physics and Department of Physics, Xiangtan University, Xiangtan 411105, Hunan, China}
\altaffiltext{4}{School of Physics and Electronic Science, Guizhou Normal University, Guiyang 550001, Guizhou, China}
\altaffiltext{5}{Key Laboratory of Dark Matter and Space Astronomy, Purple Mountain Observatory, Chinese Academy of Sciences, Nanjing 210023, Jiangsu, China}
\altaffiltext{6}{Institut f$\rm{\ddot{u}}$r Astronomie und Astrophysik, Kepler Center for Astro and Particle Physics, Eberhard Karls Universit$\rm{\ddot{a}}$t, Sand 1, 72076 T$\rm{\ddot{u}}$bingen, Germany}
\altaffiltext{7}{Department of Astronomy, Beijing Normal University, Beijing 100875, China}

\email{jiasm@ihep.ac.cn, zhangsn@ihep.ac.cn}

\begin{abstract}

  We present a detailed spectral-timing analysis of the Kilohertz quasi-periodic oscillations (kHz QPOs) in Sco X-1 using the data of Rossi X-ray Timing Explorer ($RXTE$) and the Hard X-ray Modulation Telescope ($Insight$-HXMT). The energy band with detectable kHz QPOs is studied for the first time: on the horizontal branch, it is $\sim$ 6.89--24.01 keV and $\sim$ 8.68--21.78 keV for the upper and lower kHz QPOs detected by $RXTE$, and $\sim$ 9--27.5 keV for the upper kHz QPOs by $Insight$-HXMT; on the lower normal branch, the energy band is narrower. The fractional root mean square (rms) of the kHz QPOs increases with energy at lower energy, reaches a plateau at about 16 keV and 20 keV for the lower and upper peaks, and then levels off though with a large uncertainty. The simulation of the deadtime effect of $RXTE$/PCA shows that the deadtime does not affect much the search of the kHz QPOs but makes the rms amplitude underestimated. No significant QPO is detected below $\sim$ 6 keV as shown by the $RXTE$ data, implying that the kHz QPOs do not originate from the black body emission of the accretion disk and neutron star surface. In addition, with the combined analysis of the energy spectra and the absolute rms spectra of kHz QPOs, we suggest that the kHz QPOs in Sco X-1 originate from the Comptonization of the inner part of the transition layer, where the rotation sets the frequency and the inward bulk motion makes the spectrum harder.

\end{abstract}


\section{Introduction} \label{sec:intro}

Kilohertz quasi-periodic oscillations (kHz QPOs) are the highest-frequency oscillations observed in nearly all bright neutron star low-mass X-ray binaries (LMXBs). Such oscillations were first detected in the brightest neutron star LMXB Sco X-1, soon after the launch of the Rossi X-Ray Timing Explorer ($RXTE$; \citealt{Strohmayer+1996}, \citealt{Klis+1996}, \citealt{Bradt+1993}) in 1995. Usually, kHz QPOs appear in pairs, and the twin kHz QPOs are defined as the lower and upper kHz QPOs according to their frequencies, in the range of $\sim$ 200-1200 Hz (\citealt{Klis+2006}). It is suggested that kHz QPOs in LMXBs reflect the Keplerian orbital motion at some preferred radius in the accretion disk around a neutron star (\citealt{Miller+1998}, \citealt{Cui+2000}) or represent the luminosity modulation taking place on the neutron star surface (\citealt{Gilfanov+2003}, \citealt{Gilfanov+2005}), and thus, kHz QPOs provide a possible window to study the dense matter and the strong gravity near a neutron star (\citealt{Psaltis+2008}, \citealt{Troyer+2018}). Several theoretical models on the origin of kHz QPOs have been put forward (\citealt{Miller+1998}, \citealt{Stella+1998}, \citealt{Titarchuk+2003}, \citealt{Li+2005}, \citealt{Mukhopadhyay+2009}, \citealt{Shi+2009}), but none of them can satisfactorily explain all the properties of kHz QPOs.

Sco X-1, the brightest known persistent X-ray source (\citealt{Giacconi+etal+1962}), is a $Z$ source (\citealt{Hasinger+Klis+1989})
showing a typical $Z$--shaped track on its hardness--intensity diagram (HID; \citealt{Hasinger+Klis+1989}) with three branches: the horizontal branch (HB), the normal branch (NB) and the flaring branch (FB), different from the low luminosity $Atoll$ sources with three main states: the island, lower banana and upper banana states. From this source, \cite{Klis+1996} reported the discovery of twin kHz QPOs with two peaks of $\sim$ 1100 Hz and $\sim$ 800 Hz in the energy band of 2--20 keV based on $RXTE$ data, and the frequency increases from 1050 Hz to 1130 Hz along the $Z$--track from top NB to bottom NB. A positive correlation of the frequencies between the horizontal branch oscillations (HBOs) and the upper kHz QPOs was found by \cite{Klis+1997}. The twin kHz QPOs in Sco X-1 usually present an upper peak of $\sim$ 820--1150 Hz and a lower peak of $\sim$ 540--860 Hz (\citealt{Yu+2001}, \citealt{Mendez+Klis+2000} and \citealt{Belloni+2005}). With the data of the Hard X-ray Modulation Telescope ($Insight$-HXMT; \citealt{Zhang+etal+2018}, \citealt{Li+etal+2018}, \citealt{Jia+etal+2018}, \citealt{Zhang+etal+2020}), \cite{Jia+2020} detected $\sim$ 800 Hz QPOs in Sco X-1 above 20 keV, which was the first decisive detection of the kHz QPOs in such high energy band.

Spectral-timing analysis provides an insight of how QPO properties evolve with energy and source states (\citealt{Salvo+etal+2003}, \citealt{Altamirano+etal+2008}, \citealt{Troyer+2018}), i.e., the dependence of the QPO properties (such as the centroid frequency, fractional rms and time lag) on energy or position on HIDs. It has been found that the fractional rms of kHz QPOs increases steadily with energy from $\sim$ 3 keV to 12 keV for most neutron star LMXBs (\citealt{Berger+1996}, \citealt{Zhang+etal+1996}, \citealt{Mendez+2001a}, \citealt{Ribeiro+etal+2019}). \cite{Troyer+2018} studied 14 $Atoll$ sources with the available $RXTE$ data in the energy band of 3--20 keV and found that the fractional rms of the lower kHz QPOs generally increases with energy and levels off around 15 keV, while the fractional rms of the upper kHz QPOs is less constrained at all energies. However, most of the sources with kHz QPOs studied in different energy bands are low luminosity $Atoll$ sources, while for the high luminosity $Z$ sources it is hard to observe with high time resolution in multi energy bands by $RXTE$, due to its limited onboard storage volume. So far, the energy dependence of kHz QPOs in Sco X-1 has not been studied in details, and the upper and lower limits of the energy band with detectable kHz QPOs have not been investigated yet.

$Insight$-HXMT is China's first X-ray astronomy satellite (\citealt{Zhang+etal+2020}) launched on June 15th, 2017, which carries three main instruments: the High Energy X ray Telescope (HE: 20--250 keV, geometrical area of 5000 cm$^{2}$ and time resolution of $\sim$ 2 $\mu$s; \citealt{Liu+2020}); the Medium Energy X-ray Telescope (ME: 5-30 keV, geometrical area of 952 cm$^{2}$ and time resolution of $\sim$ 276 $\mu$s; \citealt{Cao+2020}); and the Low Energy X-ray Telescope (LE: 1-15keV, geometrical area of 384 cm$^{2}$ and time resolution of $\sim$ 1 ms; \citealt{Chen+2020}). With these three instruments, $Insight$-HXMT has a broad energy band, high time resolution, large detection area at hard X-rays and non-pile-up capability when observes bright sources, making it an ideal satellite for temporal and spectral studies of bright LMXBs.

In this paper, we select the proper $RXTE$ data of Sco X-1 with multi energy bands at higher energy, together with the $Insight$-HXMT data, to search for the upper and lower energy limits of kHz QPOs, to investigate the energy dependence of the QPO properties, and further to discuss the origin of kHz QPOs. We present the data selection and reduction processes in Section 2, results in Section 3, discussions in Section 4, and conclude our study in Section 5.

\section{Observation and Data Analysis} \label{sec:obs}

\subsection{Data reduction}

To investigate the energy dependence of kHz QPOs in Sco X-1, high time resolution data with sufficient spectral information are needed. We thus search in the $RXTE$ archive for the Proportional Counter Array (PCA; \citealt{Jahoda+1996}) data of Sco X-1 with both high time resolution and multi energy bands. The observation P30406 meets these requirements, which includes two sets of single binned mode data and one set of event mode data, i.e., SB\_250us\_18\_23\_2s and SB\_250us\_24\_35\_2s with a time resolution of 250 $\mu$s and single energy bands of channel 18--23 (E1: 6.89--8.68 keV) and channel 24--35 (E2: 8.68--12.99 keV) respectively, and E\_16us\_16B\_36\_1s with a time resolution of 16 $\mu$s and 16 sub-energy bands from channel 36 to 249 (E3: 12.99--60 keV). In this observation, the peak energy band of Sco X-1 was observed in two single bands, and the hard energy band with weaker emission was divided into 16 sub bands so as to provide more data points at higher energy. This observation was taken on February 27--28, 1998 with two exposures, each covered four satellite orbits with $\sim$ 2800 s of data, separated by intervals of $\sim$ 2700 s due to Earth occultation and South Atlantic Anomaly (SAA). The integrated time of the good time intervals (GTI) in P30406 is about 22 ks.

kHz QPOs of Sco X-1 were detected in the $Insight$-HXMT observation P0101328010 (\citealt{Jia+2020}) taken on August 16, 2018, corresponding to HB in the $Z$--track. We also use this observation in the current work. The data reduction is performed similar to that in \cite{Jia+2020}, except for that a more recent version of the $Insight$-HXMT Data Analysis Software (HDAS; \citealt{Li+2020}, \citealt{Guo+2020}, \citealt{Liao+2020a}, \citealt{Liao+2020b}) V2.02\footnote{http://hxmt.org/software.jhtml} is used with much stricter GTI selection. The selected effective time is about 8 ks.

\subsection{Data analysis}

\begin{figure*}
   \centering
   \includegraphics[width=0.7\textwidth, angle=0]{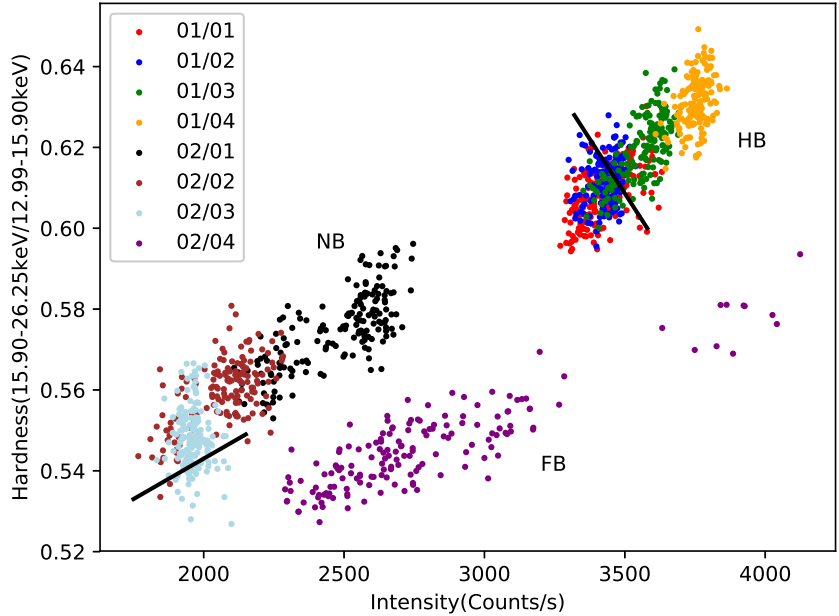}
   \caption{Hardness Intensity Diagram (HID) produced from the $RXTE/PCA$ observations of Sco X-1, P30406. Each data point corresponds to 16 s of data, and the different colors mark the observations of different orbits as denoted. The two black lines distinguish the branches of HB, NB and FB.}
   \label{Fig:HID}
   \end{figure*}

We use the 16 s data of $RXTE$/PCA to construct the HID. The hardness is defined as the ratio of the net count rates between two different energy bands, $\sim$15.90--26.25 keV and $\sim$12.99--15.90 keV\footnote{HIDs show similar shapes for different energy bands, and the $Z$ track looks more distinguishable for these two bands.}, and the intensity is defined as the net count rate in $\sim$12.99--26.25 keV. The HID is plotted in Fig.~\ref{Fig:HID}, where different colors distinguish observations performed in different orbits in P30406. The HID of Sco X-1 traces out almost a full $Z$--track, including HB, NB and FB.

We compute the power density spectra (PDS) with a time resolution of 1/4096 s using 2 s data segments (corresponding to a Nyquist frequency of 0.5--2048 Hz) for the data in P30406 observed by $RXTE$ and P0101328010 by $Insight$-HXMT. Due to the deadtime effect, the counting statistics noise no longer obeys Poisson statistics (\citealt{Zhang+etal+1995}) which will be described in Section 4.1. For inspection of kHz QPOs, we fit the 200--2000 Hz PDS with {\sc xspec} (v12.11.0), by applying an one-to-one energy-frequency conversion with a unit response (e.g. \citealt{Casella+etal+2004}). In this frequency range, a power law can represent the deadtime modified Poisson noise for $RXTE$/PCA (\citealt{Zhang+etal+1995}, \citealt{Klis+1996}, \citealt{Boutloukos+etal+2006}) and $Insight$-HXMT/HE (\citealt{Jia+2020}); while for $Insight$-HXMT/ME, the deadtime effect is significant at higher frequency ($\ge$400 Hz) and can be expressed by a power law plus a Lorentzian (\citealt{Jia+2020}). Therefore, a three-component model is applied to fit the PDS of $RXTE$/PCA and $Insight$-HXMT/HE: a power law and two Lorentzians, where the power law is used to fit the deadtime effect, and the two Lorentzians for the upper and lower kHz QPO components. A four-component model is used to fit the PDS of $Insight$-HXMT/ME, where the power law and the additional Lorentzian are used to represent the deadtime effect. It is worth to note that, whether a power law (\citealt{Klis+1996}) or a broad sinusoid (\citealt{Wijnands+1997}) is used to represent the deadtime effect in the PDS fitting, the kHz QPO properties do not change significantly.

The following parameters are obtained from the PDS fitting: the centroid frequency $L_{\rm{c}}$, the full width at half maximum (FWHM) $L_{\rm{w}}$, and the power $L_{\rm{n}}$ normalized according to Leahy (\citealt{Leahy+1983}). The quality factor ($Q$) is defined as $Q=L_{\rm{c}}/L_{\rm{w}}$ (e.g. \citealt{Barret+etal+2005}).
The fractional rms with background corrected is calculated by rms=$\sqrt{L_{\rm{n}}/(S+B)}\times(S+B)/S$ (\citealt{Belloni+etal+1990}, \citealt{Bu+etal+2015}), where $S$ and $B$ stand for source and background count rates respectively.
The significance of the signal is defined as $L_{\rm{n}}/err_{L_{\rm{n}}}$ (e.g. \citealt{Motta+etal+2015}). The errors of the quality factor and the fractional rms are estimated with a standard error propagation (\citealt{Bevington+Robinson+2003}).

\begin{figure*}[h]
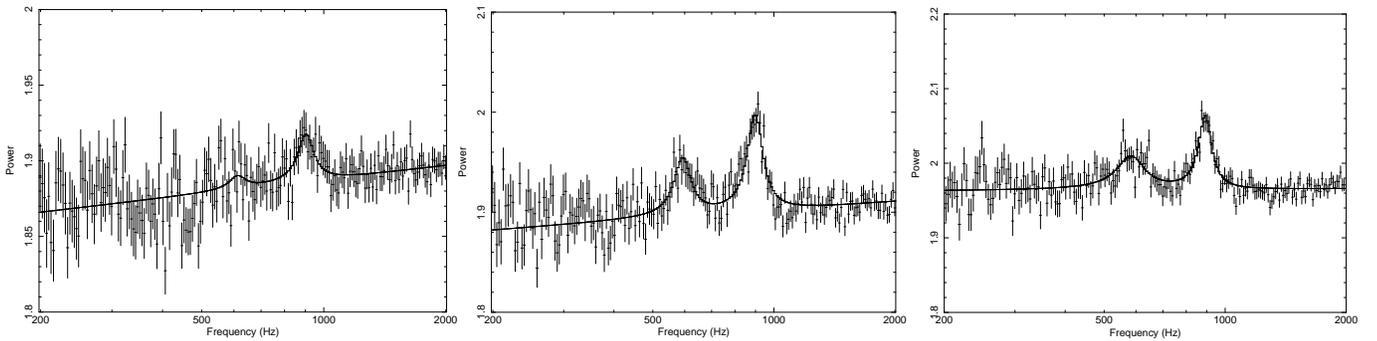

  \begin{minipage}[t]{0.33\linewidth}
  \centering
   \includegraphics[width=45mm,angle=270]{kHzQPO_0104_E1.eps}
  \end{minipage}%
  \begin{minipage}[t]{0.33\linewidth}
  \centering
   \includegraphics[width=45mm,angle=270]{kHzQPO_0104_E2.eps}
  \end{minipage}%
  \begin{minipage}[t]{0.33\linewidth}
  \centering
   \includegraphics[width=45mm,angle=270]{kHzQPO_0104_E3.eps}
  \end{minipage}%
  \caption{{The PDS of Sco X-1 fitted by a power law and two Lorentzians components for the observation P30406-01/04 of $RXTE$. Left: the kHz QPO detected in 6.89--8.68 keV. Mid: the kHz QPO detected in 8.68--12.99 keV. Right: the kHz QPO detected in 12.99--60 keV. }}
  \label{Fig:kHzQPO}
\end{figure*}

For $RXTE$/PCA data, the kHz QPOs are detected in four orbits of the first exposure and three of the second exposure, corresponding to HB and NB in the $Z$--track. Fig.~\ref{Fig:kHzQPO} shows the PDS of kHz QPOs in Sco X-1 in three energy bands of P30406-01/04. For $Insight$-HXMT data, the kHz QPOs are detected by HE and ME, which are almost the same as in \cite{Jia+2020}. The fitting results of the kHz QPOs are listed in Table~\ref{Tab:kHzQPO}, and the upper limits of the fractional rms are given if no kHz QPO is detected.

\linespread{1.8}
\begin{sidewaystable*}
\centering
\caption{The properties of kHz QPOs in Sco X-1 detected by $RXTE$ and $Insight$-HXMT respectively, where $L_{\rm{c}}$, $L_{\rm{w}}$, Q and rms represent the centroid frequency, FWHM, the quality factor and the fractional rms of kHz QPOs. For $RXTE$, E1, E2 and E3 correspond to the energy bands of 6.89--8.68 keV, 8.68--12.99 keV and 12.99--60 keV, and for $Insight$-HXMT the energy bands are 8--30 keV and 20--60 keV for ME and HE. `--' means no kHz QPO signal is detected.} \label{Tab:kHzQPO}
 \begin{tabular}{clclcccccccccc}
  \hline\noalign{\smallskip}
Satellite & Obs-ID/orbit & Energy & \multicolumn{5}{c}{Upper kHz QPO}     & \multicolumn{5}{c}{Lower kHz QPO} \\
          &              &        & $L_{\rm{c}}$ (Hz) & $L_{\rm{w}}$ (Hz) & Q & rms ($\%$) & significance
                                  & $L_{\rm{c}}$ (Hz) & $L_{\rm{w}}$ (Hz) & Q & rms ($\%$) & significance \\
  \hline\noalign{\smallskip}
$RXTE$ &P30406-01/04  & E1   & $902^{+11}_{-9}$  &  $95^{+44}_{-21}$  &  $9.5\pm3.2$   &  $2.1\pm0.3$  &  $\sim4.2\sigma$
                             & -- & -- & -- & $<$1.3   &  --  \\
       &(HB)          & E2   & $898.7^{+3.3}_{-3.2}$   &  $97.7^{+6.8}_{-6.1}$    &  $9.2\pm0.6$   &  $4.0\pm0.1$  &  $\sim13.9\sigma$
                             & $594.2^{+5.8}_{-5.4}$   &  $74^{+390}_{-10}$       &  $8^{+14}_{-8}$ & $2.5\pm0.2$  & $\sim6.1\sigma$  \\
       &              & E3   & $891.5^{+3.2}_{-2.9}$   &  $83.7^{+6.3}_{-8.2}$    &  $10.7\pm0.9$  &  $5.8\pm0.2$  &  $\sim12.9\sigma$
                             & $583.6^{+8.5}_{-8.6}$   &  $113^{+23}_{-13}$       &  $5.2\pm0.8$   &  $4.5\pm0.3$  &  $\sim6.8\sigma$  \\
       &P30406-01/03  & E1   & $918^{+11}_{-8}$        &  $83^{+17}_{-14}$        &  $11.0\pm2.1$  &  $2.0\pm0.2$  &  $\sim4.7\sigma$
                             & $686^{+15}_{-43}$       &  $57.4^{+52}_{-47}$      &  $12\pm10$     &  $1.1\pm0.3$  &  $\sim1.9\sigma$  \\
       &(HB)          & E2   & $955.1^{+4.2}_{-4.9}$   &  $94.6^{+9.4}_{-7.8}$    &  $10.1\pm0.9$  &  $3.4\pm0.2$  &  $\sim10.3\sigma$
                             & $645.0^{+7.1}_{-6.8}$   &  $106^{+14}_{-13}$       &  $6.1\pm0.8$   &  $2.9\pm0.2$  &  $\sim7.2\sigma$  \\
       &              & E3   & $934.7^{+5.0}_{-5.1}$   &  $110^{+12}_{-9}$        &  $8.6\pm0.8$   &  $5.6\pm0.3$  &  $\sim9.6\sigma$
                             & $651.8^{+7.3}_{-7.2}$   &  $95^{+19}_{-14}$        &  $6.9\pm1.2$   &  $4.3\pm0.3$  &  $\sim6.4\sigma$  \\
       &P30406-01/02  & E1   & $961^{+11}_{-14}$       &  $72^{+20}_{-15}$        &  $13.3\pm3.2$  &  $1.7\pm0.2$  &  $\sim3.9\sigma$
                             & $668.3^{+9.2}_{-9.7}$   &  $46^{+19}_{-15}$        &  $14.6\pm5.4$  &  $1.3\pm0.2$  &  $\sim3.0\sigma$  \\
       &(upper NB)    & E2   & $979.4^{+5.8}_{-5.6}$   &  $87^{+11}_{-10}$        &  $11.2\pm1.4$  &  $2.9\pm0.2$  &  $\sim7.7\sigma$
                             & $673.0^{+5.4}_{-5.4}$   &  $80^{+11}_{-10}$        &  $8.4\pm1.0$   &  $2.7\pm0.2$  &  $\sim7.8\sigma$  \\
       &              & E3   & $978.4^{+5.2}_{-5.6}$   &  $97^{+13}_{-7}$         &  $10.1\pm1.1$  &  $5.0\pm0.3$  &  $\sim7.9\sigma$
                             & $682.5^{+6.1}_{-7.1}$   &  $81^{+15}_{-10}$        &  $8.4\pm1.3$   &  $4.1\pm0.3$  &  $\sim6.7\sigma$  \\
       &P30406-01/01  & E1   & $1003.2^{+6.4}_{-5.0}$  &  $35^{+12}_{-9}$         &  $28.4\pm8.5$  &  $1.3\pm0.2$  &  $\sim3.4\sigma$
                             & $714^{+21}_{-14}$       &  $28^{+38}_{-18}$        &  $25\pm25$     &  $0.7\pm$0.3  &  $\sim1.4\sigma$  \\
       &(upper NB)    & E2   & $1009.5^{+4.0}_{-3.6}$  &  $69.5^{+7.7}_{-6.5}$    &  $14.5\pm1.5$  &  $2.9\pm0.2$  &  $\sim9.3\sigma$
                             & $721.3^{+7.6}_{-6.5}$   &  $96^{+13}_{-14}$        &  $7.5\pm1.1$   &  $2.6\pm0.2$  &  $\sim6.3\sigma$  \\
       &              & E3   & $1002.1^{+6.3}_{-5.0}$  &  $79^{+19}_{-15}$        &  $12.7\pm2.7$  &  $4.3\pm0.3$  &  $\sim6.4\sigma$
                             & $723.6^{+6.1}_{-2.8}$   &  $69\pm10$               &  $10.4\pm1.6$  &  $4.4\pm0.3$  &  $\sim8.0\sigma$  \\
       &P30406-02/01  & E1   & $1051^{+33}_{-21}$      &  60.0(fixed)             &  $\sim15.0$    &  $1.1\pm0.3$  &  $\sim1.8\sigma$
                             & -- & --  & --  & $<$0.9  &  --  \\
       &(lower NB)    & E2   & $1084.8^{+6.3}_{-6.6}$   &  $62^{+16}_{-12}$       &  $17.5\pm3.9$  &  $2.3\pm0.3$  &  $\sim 4.3\sigma$
                             & $813.7^{+9.4}_{-8.6}$    &  $38^{+26}_{-15}$       &  $21\pm11$     &  $1.4\pm0.3$  &  $\sim 2.0\sigma$  \\
       &              & E3   & $1077^{+16}_{-14}$       &  $126^{+74}_{-42}$      &  $8.6\pm3.9$   &  $4.4\pm0.7$  &  $\sim3.3\sigma$
                             & $838^{+9}_{-49}$         &  40.0(fixed)            &  $\sim21.0$    &  $2.2\pm0.4$  &  $\sim2.7\sigma$  \\
       &P30406-02/02  & E1   & $1135^{+16}_{-22}$       &  60.0(fixed)            &  $\sim16.2$    &  $1.4\pm0.3$  &  $\sim 2.6\sigma$
                             & -- & --  & --  & $<$0.9  &  --  \\
       &(lower NB)    & E2   & $1083\pm13$              &  $65^{+43}_{-23}$       &  $16.6\pm8.5$  &  $1.9\pm0.4$  &  $\sim2.7\sigma$
                             & $877^{+13}_{-15}$        &  40.0(fixed)            &  $\sim14.6$    &  $1.3\pm0.3$  &  $\sim1.9\sigma$  \\
       &              & E3   & $1137^{+13}_{-15}$       &  $58^{+54}_{-35}$       &  $19\pm15$     &  $2.9\pm0.7$  &  $\sim2.1\sigma$
                             & -- & --  & --  & $<$2.1  &  --  \\
       &P30406-02/03  & E1   & 1100.0(fixed)            &  60.0(fixed)            &  $\sim18.3$    &  $1.0\pm0.4$  &  $\sim1.2\sigma$
                             & -- & --  & --  & $<$1.0  &  --  \\
       &(lower NB)    & E2   & $1119.8^{+8.4}_{-8.1}$   & $55^{+418}_{-16}$       &  $21^{+62}_{-21}$ & $2.1\pm0.3$ & $\sim3.5\sigma$
                             & $862^{+14}_{-40}$        &  40.0(fixed)            &  $\sim21.5$    &  $1.1\pm0.3$  &  $\sim1.5\sigma$  \\
       &              & E3   & $1117.3^{+6.6}_{-6.7}$   &  $39^{+26}_{-18}$       &  $29\pm16$     &  $3.2\pm0.6$  &  $\sim2.8\sigma$
                             & $834.9^{+9.0}_{-7.8}$    &  40.0(fixed)            &  $\sim20.9$    &  $2.3\pm0.5$  &  $\sim2.1\sigma$  \\
\hline\noalign{\smallskip}
$Insight$-HXMT &P010132801001 & HE   & $804^{+10}_{-18}$ & $92^{+38}_{-26}$       & 8.8$\pm$3.0    & 13.9$\pm$0.7  & $\sim3.9\sigma$
                                     & -- & -- & -- & $<$5.7   & --   \\
               &(HB)          & ME   & $778^{+17}_{-13}$ & $68^{+19}_{-58}$       & 11.4$\pm$6.4   & 5.2$\pm$0.6   & $\sim4.1\sigma$
                                     & -- & -- & -- & $<$2.5   & --   \\
               &P010132801002 & HE   & $824^{+9}_{-11}$        & $61^{+32}_{-30}$ & 13.6$\pm$6.9   & 9.2$\pm$1.5   & $\sim3.0\sigma$
                                     & $552^{+11}_{-12}$       & $35^{+51}_{-24}$ & $16^{+23}_{-11}$& 7.8$\pm$2.5  & $\sim3.2\sigma$  \\
               &(HB)          & ME   & $840^{+8}_{-10}$        & $73^{+15}_{-63}$ & 11.5$\pm$6.2   & 4.2$\pm$0.5   & $\sim5.3\sigma$
                                     & -- & -- & -- & $<$2.2   & --   \\
  \noalign{\smallskip}\hline
\end{tabular}
\end{sidewaystable*}

We can see from Table~\ref{Tab:kHzQPO} that both the upper and lower kHz QPOs of Sco X-1 are more significant in the higher energy bands of 8.68--12.99 keV and 12.99--60 keV than in the lower energy band of 6.89--8.68 keV, and more significant on upper NB and HB (P30406-01) than on lower NB (P30406-02). It is also shown that the upper kHz QPOs are more significant than the lower kHz QPOs.

\section{Results}

\subsection{The energy limits of kHz QPOs}

Previous work usually studied the properties of kHz QPOs in Sco X-1 in one broad energy band such as 2--18 keV and 2--60 keV, without considering the energy limits of kHz QPOs. In Section 2.2 it shows that there exist significant kHz QPO signals in 12.99--60 keV of $RXTE$/PCA data, which can be divided into 16 sub-energy bands, so we use these data to search for the upper energy limit of the kHz QPOs in Sco X-1 with the method of sliding energy band as described in the next paragraph. In the following, we take the analyses of the data in P30406-01/04 and P30406-02/03 as examples, in which the source was on HB and lower NB in the $Z$--track. Since the kHz QPO signal does exist in these two observations as listed in Table~\ref{Tab:kHzQPO}, we consider it is detected in an energy band when the significance is higher than 2$\sigma$.

\begin{figure*}[h]
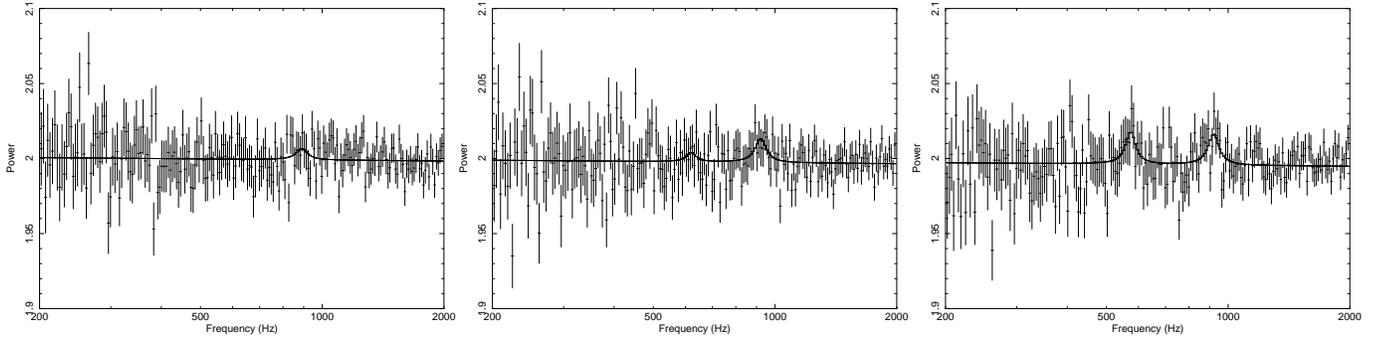

  \begin{minipage}[t]{0.33\linewidth}
  \centering
   \includegraphics[width=45mm,angle=270]{QPO_0104_65-89.eps}
  \end{minipage}%
  \begin{minipage}[t]{0.33\linewidth}
  \centering
   \includegraphics[width=45mm,angle=270]{QPO_0104_59-89.eps}
  \end{minipage}%
  \begin{minipage}[t]{0.33\linewidth}
  \centering
   \includegraphics[width=45mm,angle=270]{QPO_0104_54-89.eps}
  \end{minipage}%
  \caption{{The PDS of Sco X-1 fitted by a power law and two Lorentzians components for the observation P30406-01/04 of $RXTE$ with the kHz QPOs detected in the energy bands of 24.01--38.44 keV, 21.78--38.44 keV and 19.56--38.44 keV respectively. }}
  \label{Fig:kHzQPOEnergy}
\end{figure*}

Sco X-1 is thought to mainly radiate below 40 keV (\citealt{Duldig+1983}, \citealt{Paizis+2006}, \citealt{Revnivtsev+2014}). So, first we use the PDS in 38.44--60 keV to search for kHz QPOs, which is fitted by the three-component model as described in Section 2.2, and no signal is detected with an upper limit of $\sim$ 15\% rms in 90\% confidence level. Then, we search for the kHz QPOs with the PDS in the energy band from 38.44 keV down to the lower energies, such as 33.06--38.44 keV, 29.26--38.44 keV, 26.25--38.44 keV and so on, until the significance in which is higher than 2$\sigma$. The significance values of the kHz QPOs detected in different energy bands are listed in Table~\ref{Tab:kHzQPOEnergy}. For P30406-01/04, the significance of the upper kHz QPO is higher than 2$\sigma$ in 21.78--38.44 keV and lower than 2$\sigma$ in 24.01--38.44 keV. This means the signal above 24.01 keV is too weak to be detected, so 24.01 keV is taken as the upper energy limit of the upper kHz QPO. In the same way, the upper energy limit of the lower kHz QPO is detected as 21.78 keV. Fig.~\ref{Fig:kHzQPOEnergy} shows the PDS obtained in observation P30406-01/04 in the energy bands of 24.01--38.44 keV, 21.78--38.44 keV and 19.56--38.44 keV respectively. Similarly, for P30406-02/03, the upper energy limits of the upper and lower kHz QPOs are 19.56 keV and 15.90 keV.

\linespread{1.8}
\begin{table*}
\begin{center}
\caption{The significance of the upper and lower kHz QPOs in Sco X-1 detected by $RXTE$ in different energy bands for observations P30406-01/04 and P30406-02/03 . `--' means no kHz QPO signal is detected.
} \label{Tab:kHzQPOEnergy}
 \begin{tabular}{cccccc}
  \hline\noalign{\smallskip}
             & Energy Band      & \multicolumn{2}{c}{P30406-01/04}  & \multicolumn{2}{c}{P30406-02/03}  \\
             &                  & upper kHz QPO   & lower kHz QPO   & upper kHz QPO   & lower kHz QPO   \\
  \hline\noalign{\smallskip}
significance & 26.25--38.44 keV & --              & --              & --              & --              \\
             & 24.01--38.44 keV & $\sim1.2\sigma$ & --              & --              & --              \\
             & 21.78--38.44 keV & $\sim2.3\sigma$ & --              & --              & --              \\
             & 19.56--38.44 keV & $\sim3.7\sigma$ & $\sim3.2\sigma$ & $\sim1.3\sigma$ & --              \\
             & 18.09--38.44 keV &                 &                 & $\sim4.1\sigma$ & --              \\
             & 17.00--38.44 keV &                 &                 &                 & --              \\
             & 15.90--38.44 keV &                 &                 &                 & --              \\
             & 15.17--38.44 keV &                 &                 &                 & $\sim2.0\sigma$ \\
  \noalign{\smallskip}\hline
\end{tabular}
\end{center}
\end{table*}

In addition, the lowest energy band in P30406 is 6.89--8.68 keV, which can also be used to search for the lower energy limit of the kHz QPOs. For P30406-01/04, the significance of the upper kHz QPO in any energy band is higher than 2$\sigma$ as listed in Table~\ref{Tab:kHzQPO}, so 6.89 keV can be considered as the lower energy limit though it still needs to be confirmed with data covering lower energy, which will be done later in section 4.2. While for the lower kHz QPOs, the significance in 6.89--8.68 keV is lower than 2$\sigma$, and higher than 2$\sigma$ in 8.68--12.99 keV, so 8.68 keV is considered as the lower energy limit. Similarly, for P30406-02/03, the lower energy limit is 8.68 keV for the upper kHz QPOs and 12.99 keV for the lower kHz QPOs.

Therefore, based on the $RXTE$/PCA data, the energy range with kHz QPO in Sco X-1 is obtained as $\sim$ 6.89--24.01 keV for the upper peaks and $\sim$ 8.68--21.78 keV for the lower peaks on HB; on the lower NB, the energy ranges of the kHz QPOs are $\sim$ 8.68--19.56 keV and $\sim$ 12.99--15.90 keV for the upper and lower peaks respectively.

$Insight$-HXMT also detected the upper kHz QPOs in Sco X-1 by ME (5--30 keV) and HE (20--60 keV) in observation P0101328010 that corresponds to HB in the $Z$--track (\citealt{Jia+2020}). We search for the lower energy limit $E_{\rm{low}}$ of the kHz QPOs with the ME data from 5.0--8.0 keV by an increase of 1 keV each time till 5.0--($E_{\rm{low}}$+1) keV, in which the significance of the kHz QPO is higher than 2$\sigma$. The results are listed in Table~\ref{Tab:kHzQPOHXMT}, and it can be seen that the lower energy limit should be 9 keV. Then we search for the higher energy limit $E_{\rm{up}}$ of kHz QPOs with the HE data from 20.0--60.0 keV by an increase of 1.25 keV each time till $E_{\rm{up}}$--60.0 keV, in which the significance of the kHz QPO is lower than 2$\sigma$, as listed in Table~\ref{Tab:kHzQPOHXMT}. It means that the higher energy limit is about 27.5 keV. So, the emission energy range of the upper kHz QPOs on HB in Sco X-1 detected by $Insight$-HXMT is $\sim$ 9.0--27.5 keV. This result is roughly consistent with that of $RXTE$/PCA, but due to the smaller effective area of ME at lower energy, it is difficult to determine the lower energy limit precisely.

\linespread{1.8}
\begin{table*}
\begin{center}
\caption{The significance of the kHz QPOs in Sco X-1 detected by ME and HE of $Insight$-HXMT in different energy bands for observation P0101328010. `--' means no kHz QPO signal is detected.} \label{Tab:kHzQPOHXMT}
 \begin{tabular}{cccc}
  \hline\noalign{\smallskip}
 $Insight$-HXMT & \multicolumn{3}{c}{significance}     \\
  \hline\noalign{\smallskip}
           ME & 5--8 keV         & 5--9 keV         & 5--10 keV     \\
              & --              & $\sim1.6\sigma$ & $\sim2.6\sigma$ \\
  \hline\noalign{\smallskip}
           HE & 25--60 keV      & 26.25--60 keV   & 27.5--60 keV    \\
              & $\sim3.0\sigma$ & $\sim2.2\sigma$ & --              \\
  \noalign{\smallskip}\hline
\end{tabular}
\end{center}
\end{table*}

\subsection{The energy dependence of kHz QPOs}

\begin{figure*}[h]
  \begin{minipage}[t]{0.5\linewidth}
  \centering
   \includegraphics[width=80mm,angle=0]{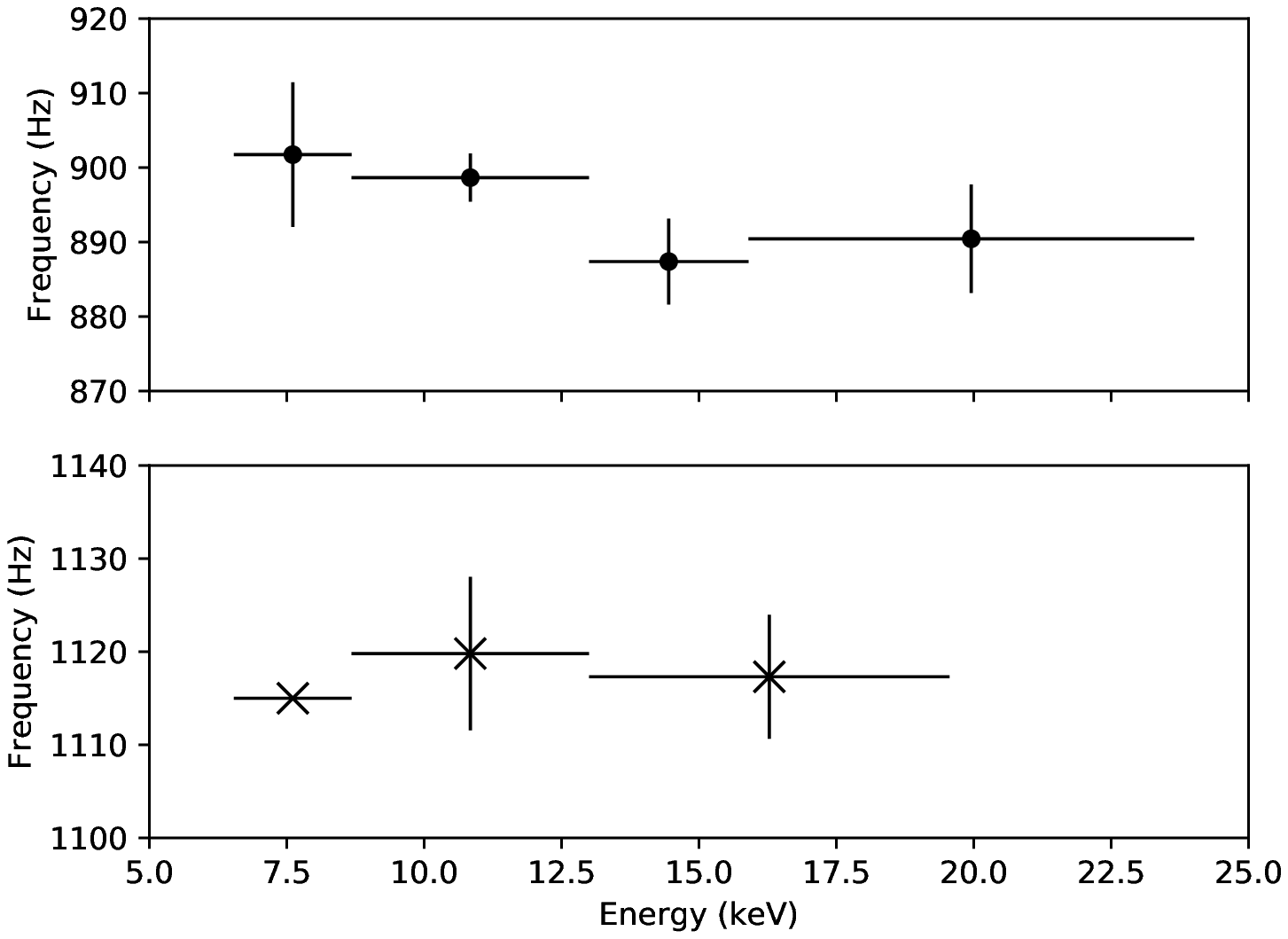}
  \end{minipage}%
  \begin{minipage}[t]{0.5\linewidth}
  \centering
   \includegraphics[width=80mm,angle=0]{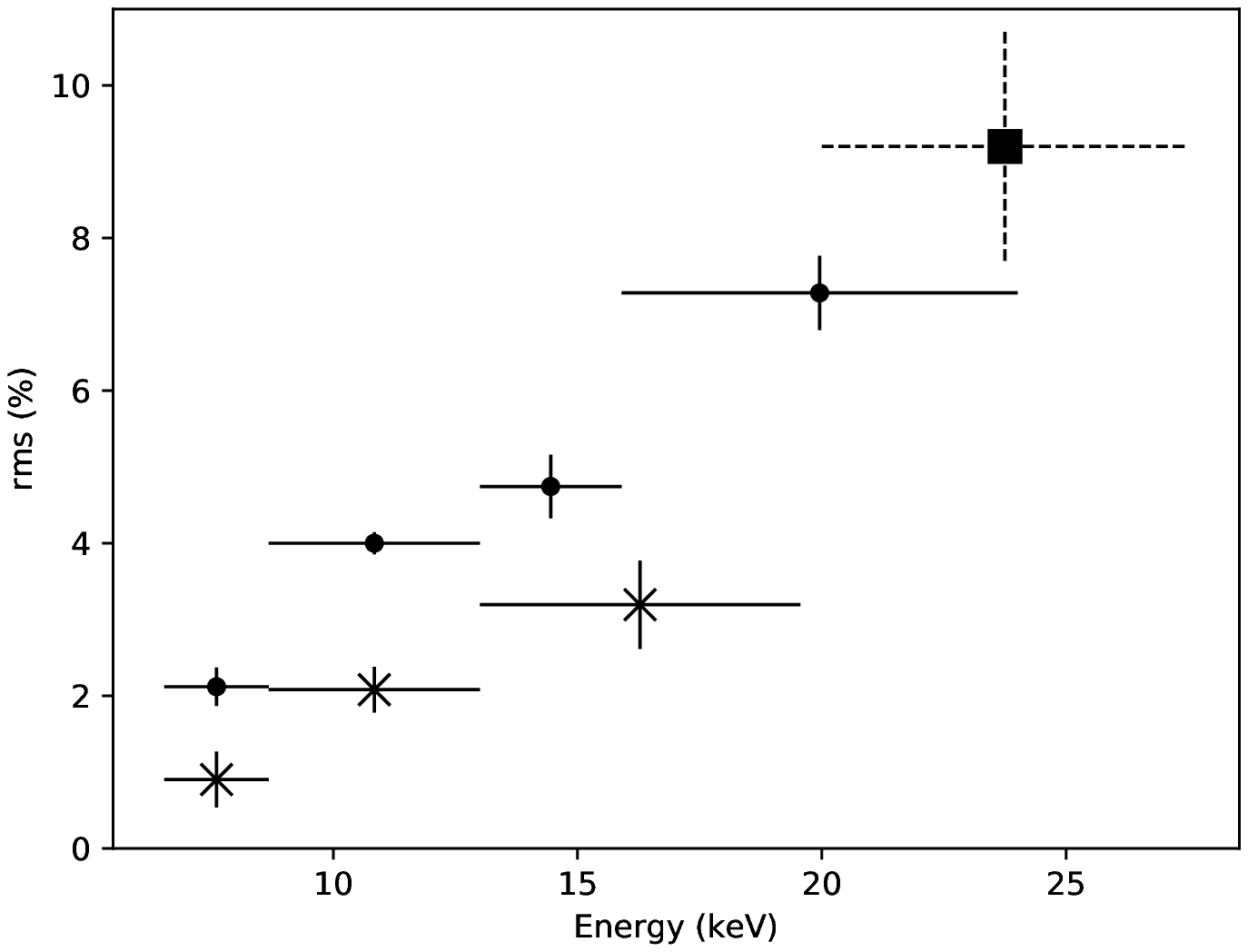}
  \end{minipage}%
  \caption{{The centroid frequencies (left) and the fractional rms (right) of kHz QPOs in Sco X-1 as a function of photon energy. The points are for the upper kHz QPOs detected in P30406-01/04 in four energy bands, the stars for the upper kHz QPOs detected in P30406-02/03 in three energy bands, and the square for the upper kHz QPO detected by $Insight$-HXMT/HE in 20--27.5 keV.}}
  \label{Fig:Edependence}
\end{figure*}

In order to investigate the energy dependence of the kHz QPOs in Sco X-1, we select four energy bands for P30406-01/04: 6.89--8.68 kev, 8.68-12.99 keV, 12.99--15.90 keV and 15.90-24.01 keV, and three energy bands for P30406-02/03: 6.89--8.68 keV, 8.68-12.99 keV and 12.99--19.56 keV, considering the significance of kHz QPO signal in each energy band and the emission energy range determined in Sect. 3.1. The PDS in each energy band is fitted with a power law plus two Lorenzians as described in Section 2.2, and the corresponding centroid frequencies and the fractional rms of the upper kHz QPOs in P30406-01/04 and P30406-02/03 are plotted in Fig.~\ref{Fig:Edependence}. It seems that the centroid frequencies of the upper kHz QPOs do not vary with energy, while the fractional rms increases steadily with energy. This is consistent with the results of $Insight$-HXMT (\citealt{Jia+2020}). In addition, the lower kHz QPOs have almost the same trend, except that they are not significant enough in some energy bands. For comparison, we plot the kHz QPO signal detected by $Insight$-HXMT/HE as the square in the right panel of Fig.~\ref{Fig:Edependence}, which presents a higher fractional rms $\sim9.2\pm1.5$\% in the higher energy band, 20--27.5 keV.

Many works have studied the energy dependence of the fractional rms of kHz QPOs in LMXBs (\citealt{Gilfanov+2003}, \citealt{Troyer+2018}, \citealt{Ribeiro+etal+2019}), but there is no consensus yet especially in the higher energy band. In Fig.~\ref{Fig:Edependence}, there are only 3-4 points, and we can not identify if there exists a break at higher energy. We thus use the overlapping energy method, like the time segment overlapping method used in the studies for the epoch and properties of the first glitch in the young pulsar PSR B0540-69 (\citealt{Ferdman+2015}), to study the energy dependence of the fractional rms.

\begin{figure*}
   \centering
   \includegraphics[width=0.8\textwidth, angle=0]{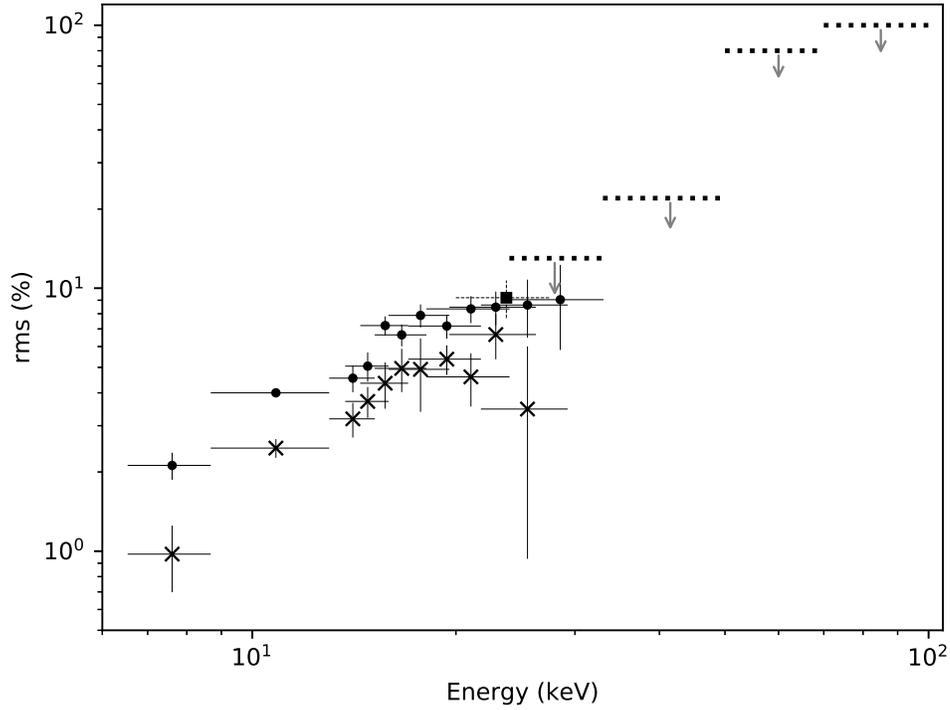}
   \caption{The fractional rms of kHz QPOs in Sco X-1 as a function of photon energy in P30406-01/04. The points are for the upper kHz QPOs, the stars for the lower kHz QPOs, the square for the upper kHz QPO detected by $Insight$-HXMT, and the arrows and the dotted lines are the rms upper limits derived by $Insight$-HXMT simulations.}
   \label{Fig:rms_overlap}
   \end{figure*}

The event mode data in P30406-01/04 are originally divided into 16 sub-energy bands from channels 36 to 249 (12.99--60 keV). Considering the photon counts, we define the new energy bands as each with three sub-energy bands of the 16 bands with overlapping energy, such as 1--3, 2--4, 3--5 and so on. The two single binned mode data with channels 18--23 (6.89--8.68 keV) and channels 24--35 (8.68--12.99 keV) are also used. Then we fit the PDS in each new energy band with the three-component model as described in Section 2.2 and derive the background corrected fractional rms of the upper and lower kHz QPOs. It shows as in Fig.~\ref{Fig:rms_overlap} that the fractional rms increases with energy and reaches a plateau at about 20 keV and 16 keV for the upper (points) and lower (stars) kHz QPOs respectively, and at higher energy it levels off without an obvious break albeit with large errors.

Fig.~\ref{Fig:rms_overlap} also shows that in Sco X-1 the fractional rms of the upper kHz QPOs are always higher than that of the lower kHz QPOs, which is consistent with the previous results in \cite{Klis+1996}, \cite{Klis+1997} and \cite{Yu+2001}. This property is similar to that of the $Z$ source GX 17+2 (\citealt{Klis+2006}) but different from that of $Atoll$ sources, in which the lower kHz QPOs are more obvious (see \citealt{Sanna+2010}, \citealt{Mukherjee+Bhattacharyya+2012}, \citealt{Peille+2015}, \citealt{Troyer+2018}). This may be one of the main differences between $Z$ sources and $Atoll$ sources.

\subsection{The variability of kHz QPOs along $Z$-track}

\begin{figure*}
   \centering
   \includegraphics[width=0.6\textwidth, angle=0]{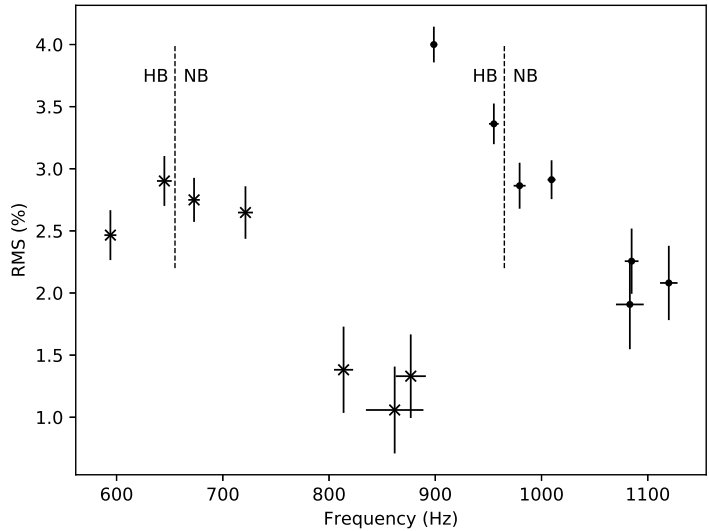}
   \caption{The variation of the centroid frequencies and fractional rms of kHz QPOs in Sco X-1 in the energy band of 8.68--12.99 keV along the $Z$--track. The points are for the upper kHz QPOs, the stars for the lower kHz QPOs, and from left to right corresponds to from HB to lower NB.}
   \label{Fig:Fre_rms}
   \end{figure*}

We use the $RXTE$/PCA data of P30406 in 8.68--12.99 keV to study the variability of kHz QPOs along the $Z$--track, as in this energy band the signals are significant for both the upper and lower kHz QPOs. We fit the PDS of the seven-orbit data with detectable kHz QPOs as presented in Table~\ref{Tab:kHzQPO}, which distribute on HB, upper NB and lower NB of the $Z$--track. The fractional rms of the upper and lower kHz QPOs as functions of the centroid frequencies are plotted in Fig.~\ref{Fig:Fre_rms}, where the points are for the upper kHz QPOs and stars the lower kHz QPOs. It shows that along the $Z$--track from HB to lower NB, the centroid frequencies increase from $\sim$ 900 Hz to $\sim$ 1100 Hz for the upper kHz QPOs and from $\sim$ 600 Hz to $\sim$ 900 Hz for the lower kHz QPOs, while the fractional rms decreases from 4\% to 2\% and from 3\% to 1\% for the upper and lower kHz QPOs respectively. These results are consistent with those found in Sco X-1 by \cite{Klis+1996}, \cite{Klis+1997} and \cite{Bradshaw+2008}, and this phenomenon is also detected in GX 17+2 (\citealt{Wijnands+1997}), 4U 1608-52 (\citealt{Berger+1996}), 4U 1728-34 and Aquila X-1 (\citealt{Mendez+2001b}).

\section{DISCUSSION}

In the previous section, we have detected the kHz QPOs in Sco X-1 and searched for the emission energy band with detectable kHz QPOs for the first time. The full coverage of the $Z$--track and the broad energy band of $RXTE$ and $Insight$-HXMT provide us an opportunity to study the evolution of kHz QPO with energy band and spectral states: the centroid frequency of kHz QPO is energy independent, while its fractional rms increases with energy and levels off at about 20 keV; along the $Z$--track from HB to lower NB, the centroid frequency increases and the rms decreases. In the following, we will compare our results with the previous ones, discuss about the deadtime effect on kHz QPO properties, study the energy dependence of the fractional rms, and explore the origin of kHz QPOs.

\subsection{The deadtime effect on kHz QPOs}

Due to the deadtime effect of the instruments used, the power spectra deviate from the Poisson statistics (\citealt{Zhang+etal+1995}, \citealt{Boutloukos+etal+2006}), which could influence the determination of QPO properties. For the high-frequency analysis of a bright source such as Sco X-1, the deadtime effect could be more significant. In order to estimate this effect, we simulate the observations of $RXTE$/PCA for Sco X-1 with an exposure time of 2000 s, count rate of 20000 cnts/s, QPO centroid frequency of 900 Hz, FWHM of 90 Hz and fractional rms of 10\%, in no-deadtime and deadtime-affected cases respectively. The deadtime effect is simulated by discarding the events whose recorded arrival times are less than 10 $\mu$s from their respective previous events, where 10 $\mu$s is the deadtime of $RXTE$/PCA.

\begin{figure*}[h]
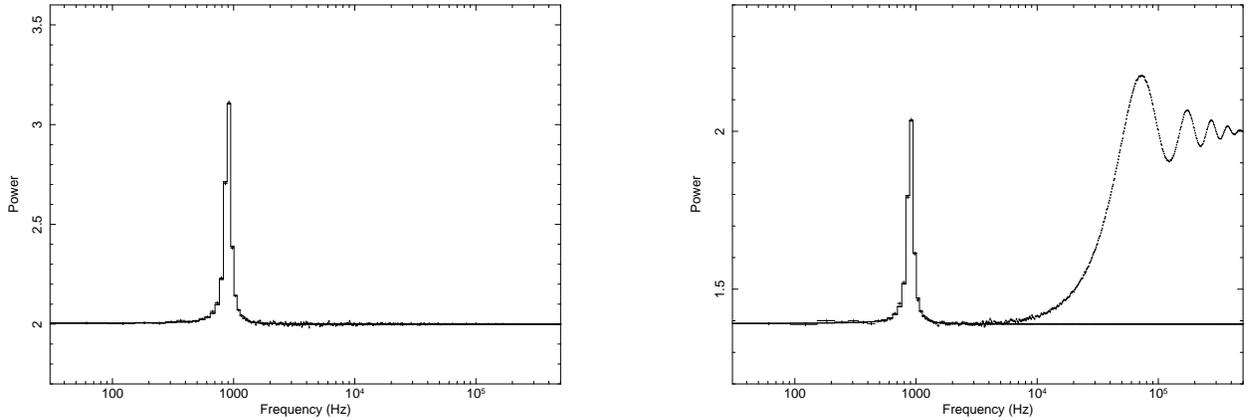

  \begin{minipage}[t]{0.5\linewidth}
  \centering
   \includegraphics[width=55mm,angle=-90]{PDS_nodeadtime.eps}
  \end{minipage}%
  \begin{minipage}[t]{0.5\linewidth}
  \centering
   \includegraphics[width=55mm,angle=-90]{PDS_deadtime.eps}
  \end{minipage}%
  \caption{{The PDS of the simulated observations of $RXTE$/PCA for Sco X-1 in no-deadtime (left panel) and deadtime-affected (right panel) cases, and the best-fit line (the solid line) with the model of a power law plus a Lorentzian in 200--2000 Hz.}}
  \label{Fig:deadtime}
\end{figure*}

The power spectra with and without deadtime are plotted in Fig.~\ref{Fig:deadtime}. It shows that the average Poisson level is $\sim$ 2 for the PDS without deadtime, and when the deadtime is included, the average Poisson level significantly deviates from $\sim$ 2 and presents a `wavy' shape at higher frequency ($>10000$ Hz), while the PDS shape at frequency less than 5000 Hz seems to be flat. This means that the `wavy' will not affect the search of kHz QPO signal. To further quantify the deadtime effect on the QPO rms, we fit the PDS in 200--2000 Hz with a power law plus a Lorentzian model as described in Section 2.2, where the power law component is used to fit the deadtime effect and the Lorentzian for the kHz QPO. The fitting results are plotted as the solid lines in Fig.~\ref{Fig:deadtime} and listed in Table~\ref{Tab:deadtime_fitting}. It shows that the fractional rms are 8.4\% and 9.5\% for cases with and without deadtime respectively, indicating that the kHz QPO rms amplitude in Sco X-1 may be underestimated by $\sim$ 12\% due to the deadtime effect. We note here that, our simulation is for the idealized case, in which only a quasi-periodical signal with fixed centroid frequency is added to the light curve, while in the real data, there are many non-periodical variations and the QPO centroid frequency evolves with time. Thus, in the simulation the kHz QPO signal is more significant than that from the real $RXTE$/PCA data; while in the real data, the non-periodical signal is superimposed on the Poisson noise, making the average noise level to have a slope rather than a constant as in the simulation.

The deadtime correction method as described in $RXTE$ Cook Book\footnote{https://heasarc.gsfc.nasa.gov/docs/xte/recipes/pca\_deadtime.html} shows that the deadtime effect is energy independent. For$RXTE$/PCA, the data in different energy bands are readout by the same set of electronics. The event of any energy causes deadtime during the readout, and the amount of deadtime is the same for all the events, so the deadtime effect is energy independent. Accordingly, the deadtime effect affects the value of the kHz QPO rms, but not the trend with energy as in Fig.~\ref{Fig:Edependence} and Fig.~\ref{Fig:rms_overlap}. Moreover, the deadtime of $Insight$-HXMT/HE is $\sim$ 2 $\mu$s, which will have less effect on the kHz QPO study than that of $RXTE$/PCA.

\linespread{1.8}
\begin{table*}
\begin{center}
\caption{The fitting results of the PDS in 200--2000 Hz with the model of a power law plus a Lorentzian for no-deadtime and deadtime-affected cases.} \label{Tab:deadtime_fitting}
 \begin{tabular}{llllllll}
  \hline\noalign{\smallskip}
                & \multicolumn{2}{c}{power law} & \multicolumn{3}{c}{Lorentzian} &  &  \\
                & index       & norm            & $L_{\rm{c}}$ (Hz) & $L_{\rm{w}}$ (Hz) & norm & $\chi$$^{2}$/dof   & rms(\%)     \\
  \hline\noalign{\smallskip}
    deadtime    & 0.0001$\pm$0.0007   & 1.391$\pm$0.006 & 900.9$\pm$0.8 & 92.2$\pm$2.3 & 109.7$\pm$1.6 & 36.6/48 & 8.4$\pm$0.1 \\
    no deadtime & 0.00003$\pm$0.00005 & 2.000$\pm$0.002 & 900.4$\pm$0.5 & 93.6$\pm$2.0 & 191.1$\pm$2.2 & 38.5/48 & 9.5$\pm$0.1 \\
  \noalign{\smallskip}\hline
\end{tabular}
\end{center}
\end{table*}

\subsection{The lower and upper energy limits of kHz QPOs}

Our results in Section 3.1 show that the upper kHz QPOs in Sco X-1 are significant only in $\sim$ 6.89--24.01 keV for $RXTE$/PCA and $\sim$ 9--27.5 keV for $Insight$-HXMT. In comparison, \cite{Troyer+2018} investigated the energy dependence of the fractional rms of kHz QPOs from 3 keV to 20 keV in 14 LMXBs, which show significant kHz QPOs at around 3 keV. Since there is no observation below 6.89 keV in the $RXTE$ data P30406, and $Insight$-HXMT/ME is not sensitive enough below $\sim$ 8 keV (\citealt{Cao+2020}), further study is needed to verify if the kHz QPO signal in Sco X-1 exists at lower energy.

\begin{figure*}[h]
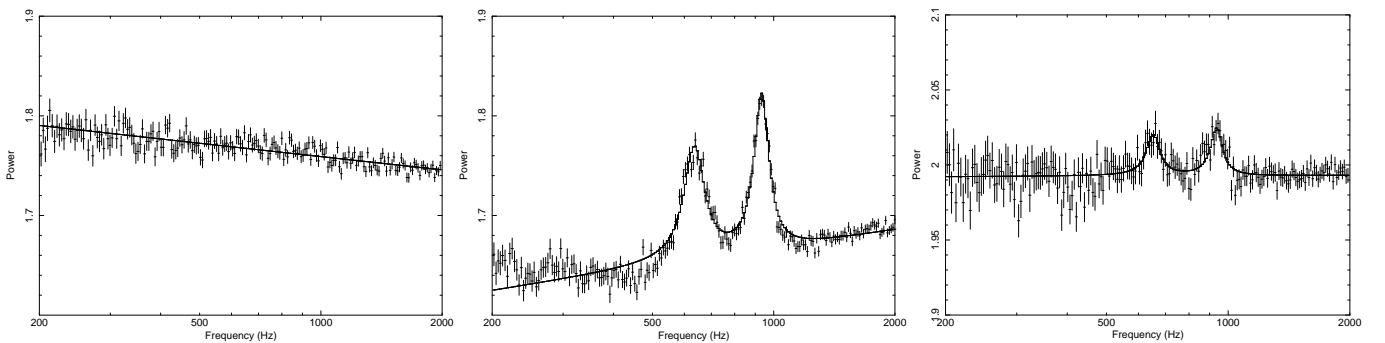

  \begin{minipage}[t]{0.33\linewidth}
  \centering
   \includegraphics[width=45mm,angle=270]{R_QPO_0-13.eps}
  \end{minipage}%
  \begin{minipage}[t]{0.33\linewidth}
  \centering
   \includegraphics[width=45mm,angle=270]{R_QPO_14-49.eps}
  \end{minipage}%
  \begin{minipage}[t]{0.33\linewidth}
  \centering
   \includegraphics[width=45mm,angle=270]{R_QPO_50-249.eps}
  \end{minipage}%
  \caption{The PDS of Sco X-1 fitted by a power law and two Lorentzians components for the observation P30035 of $RXTE$ in the energy bands of 1.94--5.12 keV, 5.12--18.09 keV and 18.09--60 keV respectively. }
  \label{Fig:QPO_hard}
\end{figure*}

Here, we reanalyse another set of $RXTE$/PCA data studied by \cite{Mendez+Klis+2000}, P30035, observed on July 4, 1998. Two observation mode data with high time resolution of 250 $\mu$s in P30035 can be used for the kHz QPO studies: one is the single bin mode data in channels 50 to 249 (18.09--60 keV), and the other is the bin mode data in channels 0 to 49 (1.94--18.09 keV) which can be divided into two sub-bands of channels 0--13 (1.94--5.12 keV) and channels 14--49 (5.12--18.09 keV). In order to study the kHz QPOs in lower energy band, we focus on the separated energy bands, 1.94--5.12 keV, 5.12--18.09 keV and 18.09--60 keV. The PDS of each energy band is fitted with the three-component model as described in Section 2.2. Shown as in Fig.~\ref{Fig:QPO_hard}, the kHz QPO signal is not detected in 1.94--5.12 keV with an upper limit of $\sim$ 0.7\% rms in 90\% confidence level; the signal is very strong in 5.12--18.09 keV with the significance of $\sim$ 20$\sigma$; while in 18.09-60 keV the significance values of the kHz QPOs are $\sim$ 3.3$\sigma$ and $\sim$ 4.8$\sigma$ for the lower and upper peaks, which are similar to the results of $Insight$-HXMT (\citealt{Jia+2020}) in which the significance is $\sim$ 3--5$\sigma$. Combining the results of $RXTE$ and $Insight$-HXMT, we can conclude that most probably there exists no kHz QPO signal below $\sim$ 6 keV, and the lower energy limit of kHz QPOs in Sco X-1 is about 6 keV. Since the black body temperature of the accretion disk is typically lower than $\sim$ 0.7 keV in LMXBs (see \citealt{Barret+2001}, \citealt{Titarchuk+2014}), our result also suggests that the kHz QPO signal in Sco X-1 does not come from the black body emission of the accretion disk and neutron star surface.

The upper energy limits of the kHz QPOs in Sco X-1 detected by $RXTE$/PCA and $Insight$-HXMT/HE are about 24.0 keV and 27.5 keV respectively as presented in Section 3.1, but there exist large uncertainties. In order to investigate whether the large errors or even non-detection of the fractional rms of kHz QPOs at higher energy is a consequence of weaker signal or lower sensitivity of the instrument, we simulate the observations of Sco X-1 by $RXTE$/PCA and $Insight$-HXMT/HE in 24--33 keV with an exposure time of 2000 s, and the deadtime effects are considered with the same method as in Section 4.1. The simulation results listed in Table~\ref{Tab:simulation} show that if the fractional rms is set to about 10\%, neither $RXTE$/PCA nor $Insight$-HXMT/HE can detect the kHz QPO signal with high enough significance. When the fractional rms reaches $\sim$ 14\% and $\sim$ 26\%, the signal can be detected with a significance higher than 3$\sigma$ by $Insight$-HXMT/HE  and $RXTE$/PCA, respectively. This implies that the fractional rms of kHz QPOs do not increase continuously at higher energy with the same rate as in $\sim$ 6--20 keV, otherwise it should have been detected by $Insight$-HXMT/HE. Our simulations further give the upper limits of the fractional rms to 100 keV for the kHz QPOs in Sco X-1, as shown in Table~\ref{Tab:simulation} and Fig.~\ref{Fig:rms_overlap}. It indicates that at energy much higher than 30 keV, $Insight$-HXMT can not give a strict constraint on kHz QPOs, due to mainly the increase of the background and decrease of the source flux. Moreover, our simulations show that when the observation time increases to $\sim$ 8000 s, $Insight$-HXMT/HE can detect the kHz QPO signal with fractional rms of $\sim$ 10\% in 24--33 keV, at significance level higher than 3$\sigma$. However, since the frequencies of kHz QPOs evolve rapidly along the $Z$--track, it is hard to accumulate 8000 s of observation data at a specific QPO frequency.

Therefore, the upper energy limits of the kHz QPOs detected by $RXTE$/PCA and $Insight$-HXMT/HE still have considerable uncertainties. Similar results were also reported by \cite{Ribeiro+etal+2019}. They found that, due to the limited capability of RXTE/PCA at high energies, the energy dependence of the rms amplitude of the upper kHz QPO in 4U 1636-53 can not be well determined. We expect future X-ray missions with larger effective area at high energy, such as the enhanced X--ray Timing and Polarimetry mission (eXTP; \citealt{Zhang+2019}), to give a better restriction on the fractional rms distribution and the upper energy limits of kHz QPOs in Sco X-1.

\linespread{1.8}
\begin{table*}
\begin{center}
\caption{The results of the simulated observation of $RXTE$ and $Insight$-HXMT in different energy bands with an exposure time of 2000 s.} \label{Tab:simulation}
 \begin{tabular}{llllllllccc}
  \hline\noalign{\smallskip}
                & \multicolumn{3}{c}{$RXTE$} & & \multicolumn{6}{c}{$Insight$-HXMT} \\
                & \multicolumn{3}{c}{24--33 keV} & & \multicolumn{3}{c}{24--33 keV} & 33--50 keV & 50--70 keV & 70--100 keV \\
  \hline\noalign{\smallskip}
   rms          &10\%        &24\%        &26\%        & &10\%        &13\%        &14\%        &22\%        &80\%        &100\%       \\
   significance &1.3$\sigma$ &2.1$\sigma$ &3.1$\sigma$ & &1.5$\sigma$ &2.3$\sigma$ &3.3$\sigma$ &3.1$\sigma$ &3.3$\sigma$ &2.9$\sigma$ \\
   \noalign{\smallskip}\hline
\end{tabular}
\end{center}
\end{table*}

\subsection{Constraints on the origin of kHz QPOs}

A common picture presents that in a neutron star LMXB a Keplerian disk is connected to the neutron star through a transition layer, and the accretion onto the neutron star takes place when the material passes through the two major regions: the accretion disk and the transition layer (\citealt{Gilfanov+2005}). Accordingly, \cite{Titarchuk+2014} found that a two--Comptb model, wab*(Comptb+Comptb+Gauss), can express the emission process in the neutron star LMXBs and give satisfactory fits for all the available spectra of Sco X-1.

The Comptb model (\citealt{Farinelli+2008}) describes an emergent spectrum as convolution of an input seed black body spectrum with a Compotonization Green function, where both the thermal and dynamical (i.e., bulk) effects can be taken into account. The free parameters in Comptb are: seed photon temperature $kT_{\rm{s}}$, electron temperature $kT_{\rm{e}}$, spectral index $\alpha$, bulk parameter $\delta$, and illumination factor log$A$, where $1/(1+A$) is the fraction of the seed-photon radiation directly seen by the observer and  $A/(1+A)$ the fraction upscattered by the Compton cloud. Therefore, in the two-Comptb scenario (\citealt{Titarchuk+2014}), one Comptb accounts for the Comptonization spectrum of the disk photons scattered off by the outer part of the transition layer between the accretion disk and the neutron star, where the bulk inflow effect is negligible compared to the thermal Comptonization and the bulk parameter can be fixed as $\delta_1=0$; the other Comptb is related to the photons from the neutron star scattered off by the inner part of the transition layer, where the bulk effect $\delta_2$ should be taken into account as it is close to the neutron star; and a Gauss model is added to account  for the Fe K line. Here, we use the two-Comptb model to fit the energy spectra and the absolute rms spectrum of kHz QPOs in Sco X-1 to identify which component is more likely associated with the millisecond variable part of the X-ray emission, the kHz QPOs.

\begin{figure*}[h]
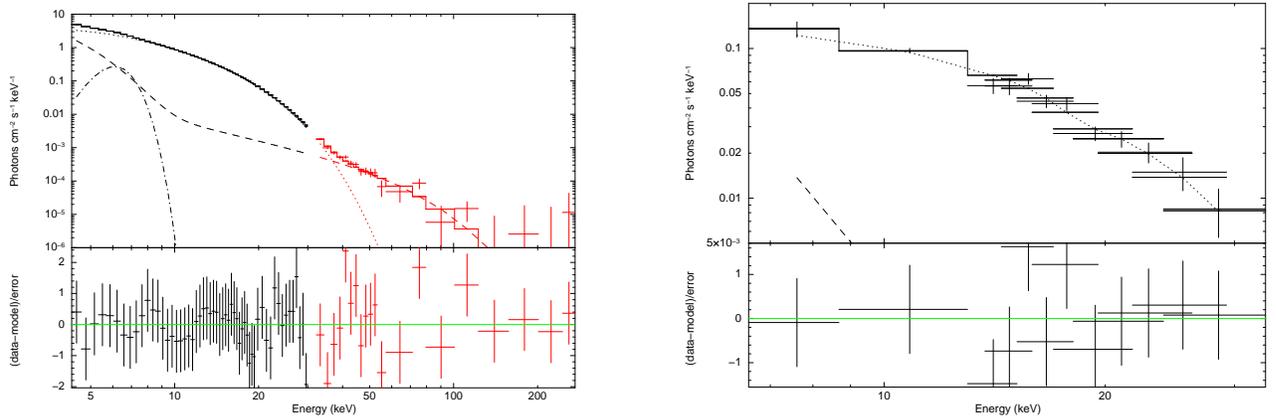

  \begin{minipage}[t]{0.5\linewidth}
  \centering
   \includegraphics[width=55mm,angle=270]{spectra_DC_color.eps}
  \end{minipage}%
  \begin{minipage}[t]{0.5\linewidth}
  \centering
   \includegraphics[width=55mm,angle=270]{spectra_rms_DC_color.eps}
  \end{minipage}%
  \caption{The energy spectrum (left panel) and absolute rms spectrum (right panel) of P30406-01/04 after dead time correction fitted by the two--Comptb model. Comptb1 is displayed by the dashed line, Comptb2 by the dotted lime, and Gauss by the dash-dotted line.}
  \label{Fig:spec}
\end{figure*}

We extract the energy spectra of $RXTE$/PCA and $RXTE$/HEXTE with the command $rex$. In this pipeline process, the deadtime effect of the HEXTE spectra has been corrected, while that of the PCA spectra need to be further dealt with. Here, we carry out the deadtime correction for the PCA spectra by calculating a new exposure time with the deadtime fraction ($DTF$) and the deadtime correction factor ($DCOR$) as described in $RXTE$ Cook Book\footnote{https://heasarc.gsfc.nasa.gov/docs/xte/recipes/pca\_deadtime.html}, where $DTF=0.69$ and $DCOR=1.44$. The energy spectra of PCA and HEXTE for the observation P30406-01/04 are plotted in the left panel of Fig.~\ref{Fig:spec}, which can be fitted well with the two-Comptb model, where $Comptb1$ (dashed line) accounts for the photons from accretion disk scattered off by the outer part of the transition layer and $Comptb2$ (dotted line) for the photons from neutron star scattered off by the inner part of the transition layer. It shows that $Comptb2$ dominates the emission below $\sim$ 35 keV, and $Comptb1$ dominates the harder emission above $\sim$ 35 keV.
The best-fit parameters are listed in Table~\ref{Tab:rms_fitting}. $\delta_2=5.6^{+4.9}_{-2.9}$ means that in the inner part of the transition layer there exists bulk effect. The best fit value of log$A_2$ is much larger than 2, indicating that the Comptonization fraction, $A/(1+A)$, approaches unity and the fraction of the black body emission from neutron star, $A/(1+A)$, approaches zero. In this case, any variation of log$A$ can not change the other fitting results significantly (e.g., \citealt{Titarchuk+2014}), and we thus fix log$A_2=2$ in the fitting process. Therefore, $Comptb2$ is dominated by the Comptonization (thermal plus bulk) component of the inner part of the transition layer, not by the black body emission of the neutron star.

We also derive the absolute rms spectrum $S_{\rm{RMS}}$ of the upper kHz QPOs with the fractional rms in Fig.~\ref{Fig:rms_overlap}, $S_{\rm{RMS}}(E)=\rm{rms}\times$$R/B$, where $B$ is the effective area of the detector and $R$ the deadtime corrected count rate which is related to the detected count rate $R_{0}$, $R=R_{0}/(1-DTF)$. In order to identify which component is more likely associated with the kHz QPO signal, we fit the absolute rms spectrum with the same two-Comptb model as we do in the energy spectral fitting. The results are also listed in Table~\ref{Tab:rms_fitting}. Fig.~\ref{Fig:spec} shows that the absolute rms spectrum is dominated by $Comptb2$ (the dotted line) that is related to the photons from neutron star and the inner part of the transition layer. Similar to that in the energy spectral analysis, log$A_2=2$ indicates that Comptb2 is dominated by the Compton emission of the inner part of the transition layer. This also suggests that the kHz QPOs originate from the Comptonization of the inner part of the transition layer. Given that the $Comptb2$ component in the absolute rms spectrum is much harder (with a smaller $\alpha_2$) than that of the energy spectra, and the bulk effect ($\delta_2=5.6^{+4.9}_{-2.9}$) can increase the cutoff energy (\citealt{Farinelli+2008}), we speculate that the kHz QPO is more related to the bulk motion of the inner part of the transition layer.

\linespread{1.8}
\begin{table*}
\begin{center}
\caption{The fitting results of the energy spectra and the absolute rms spectra of kHz QPO in Sco X-1 with the two--Comptb model.} \label{Tab:rms_fitting}
 \begin{tabular}{llll}
  \hline\noalign{\smallskip}
  \multicolumn{2}{c}{$Model$}  & Energy spectrum                    & rms spectrum \\
  \hline\noalign{\smallskip}
  Comptb$_{1}$ & $T_{\rm{s1}}$ (keV) & $0.71^{+0.13}_{-0.09}$       & = 0.71   \\
               & $\alpha_1$          & $0.99\pm0.07$                & = 0.99   \\
               & $\delta_1$          & 0 (fixed)                    & = 0      \\
               & $T_{\rm{e1}}$ (keV) & $13.0^{+3.6}_{-2.3}$         & = 13.0   \\
               & log$A_1$            & -$1.98^{+0.07}_{-0.08}$      & = -1.98  \\
               & $norm_1$            & $1.42^{+0.28}_{-0.55}$       & $0.10^{+0.48}_{-0.10}$  \\
  Comptb$_{2}$ & $T_{\rm{s2}}$ (keV) & $1.33^{+0.11}_{-0.22}$       & = 1.33   \\
               & $\alpha_2$          & $1.51^{+0.28}_{-0.33}$       & $0.54\pm0.07$ \\
               & $\delta_2$          & $5.6^{+4.9}_{-2.9}$          & = 5.6    \\
               & $T_{\rm{e2}}$ (keV) & $1.94^{+0.42}_{-0.47}$       & = 1.94   \\
               & log$A_2$            & 2.0 (fixed)                  & = 2.0    \\
               & $norm_2$            & $1.91^{+0.06}_{-0.03}$       & $0.14^{+0.02}_{-0.01}$ \\
  $\chi$$^{2}$/dof &                 & 85.02/95                     & 7.58/12  \\
  \noalign{\smallskip}\hline
\end{tabular}
\end{center}
\end{table*}

Our results set some new constraints on the kHz QPO models. Some theoretical models suggest that the kHz QPOs in LMXBs are relevant to the Keplerian orbital motion at some preferred radii in the inner accretion disk around a neutron star (\citealt{Miller+1998}, \citealt{Cui+2000}). \cite{Gilfanov+2003} and \cite{Gilfanov+2005} considered that the emission of the accretion disk is less variable on $\sim$ sec-msec time scales and the kHz QPOs represent the luminosity modulation taking place on the neutron star surface. In our work, no significant kHz QPO is detected below $\sim$ 6 keV, implying that the signal can not originate from the black body emission of the accretion disk and the neutron star surface. With the combined analysis of the energy and the absolute rms spectra, it is found that the kHz QPOs come mostly from the Compton emission of the inner part of the transition layer instead of from the neutron star.

\cite{Li+2005} presented a plausible mechanism for the production of kHz QPOs in neutron star LMXBs, in which the stellar magnetic fields truncate the disk, resulting in a sub-Keplerian boundary layer, and the rotation frequency (sub-Keplerian) of the boundary layer leads to the upper kHz QPOs. Their model explains the origin of the frequency of kHz QPO, but the discussion about spectrum is absent. Our results show that the kHz QPO presents a much harder spectrum, which may be related to the Comptonization of the soft photons from the neutron star by the bulk motion of the inner part of the transition layer (e.g., \citealt{Niedzwiecki+2006}, \citealt{Farinelli+2008}). A speculated physical scenario based on \cite{Li+2005} and our analysis could be: the interaction between the neutron star magnetic field and the surrounding accretion disk results in a sub-Keplerian transition layer; the sub-Keplerian rotation of the inner part of the transition layer leads to the kHz QPO and the inward bulk motion explains the presence of high-energy photons. In short, we suggest that the kHz QPOs in Sco X-1 originate from the Comptonization of the inner part of the transition layer, where the rotation sets the frequency and the inward bulk motion makes the spectrum harder. This scenario is depicted in Fig. 10, adapted from the Fig 2 in \cite{Farinelli+2008}.

\begin{figure*}[h]
  \begin{minipage}[t]{0.65\linewidth}
  \centering
   \includegraphics[width=115mm,angle=0]{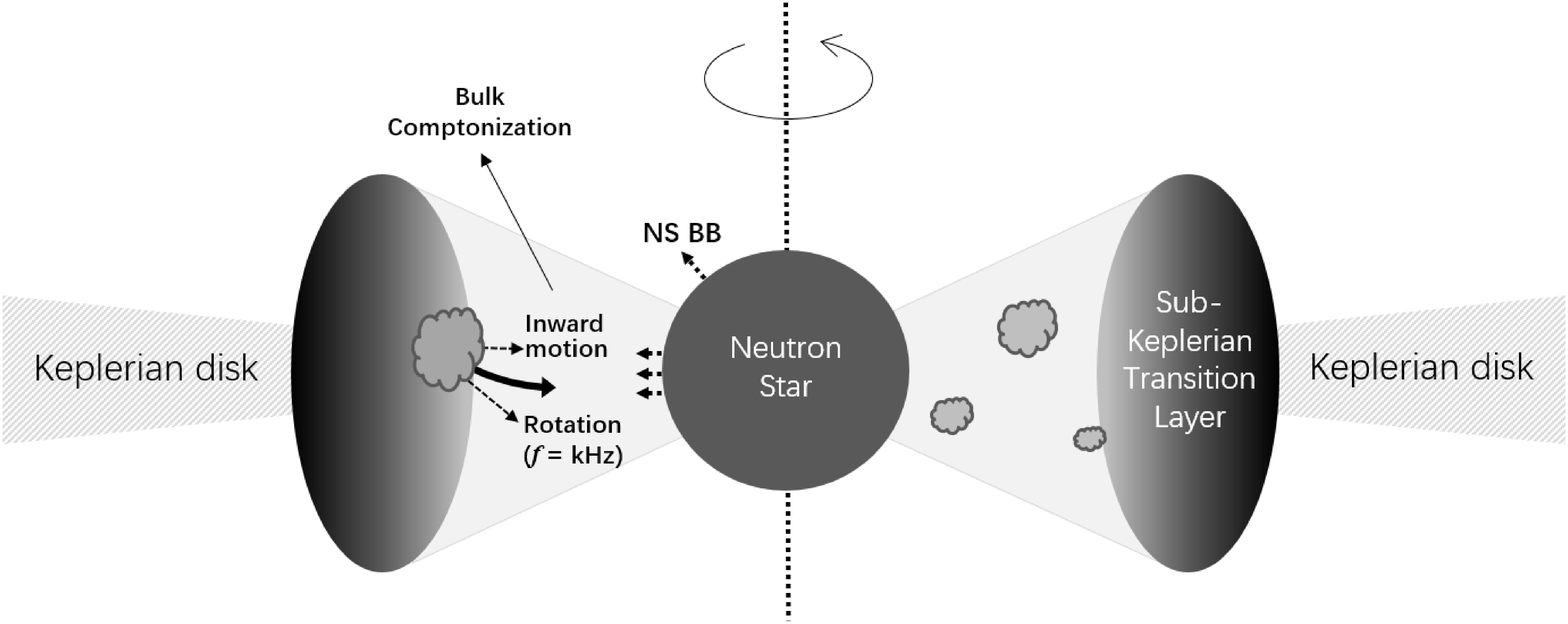}
  \end{minipage}%
  \begin{minipage}[t]{0.35\linewidth}
  \centering
   \includegraphics[width=55mm,angle=0]{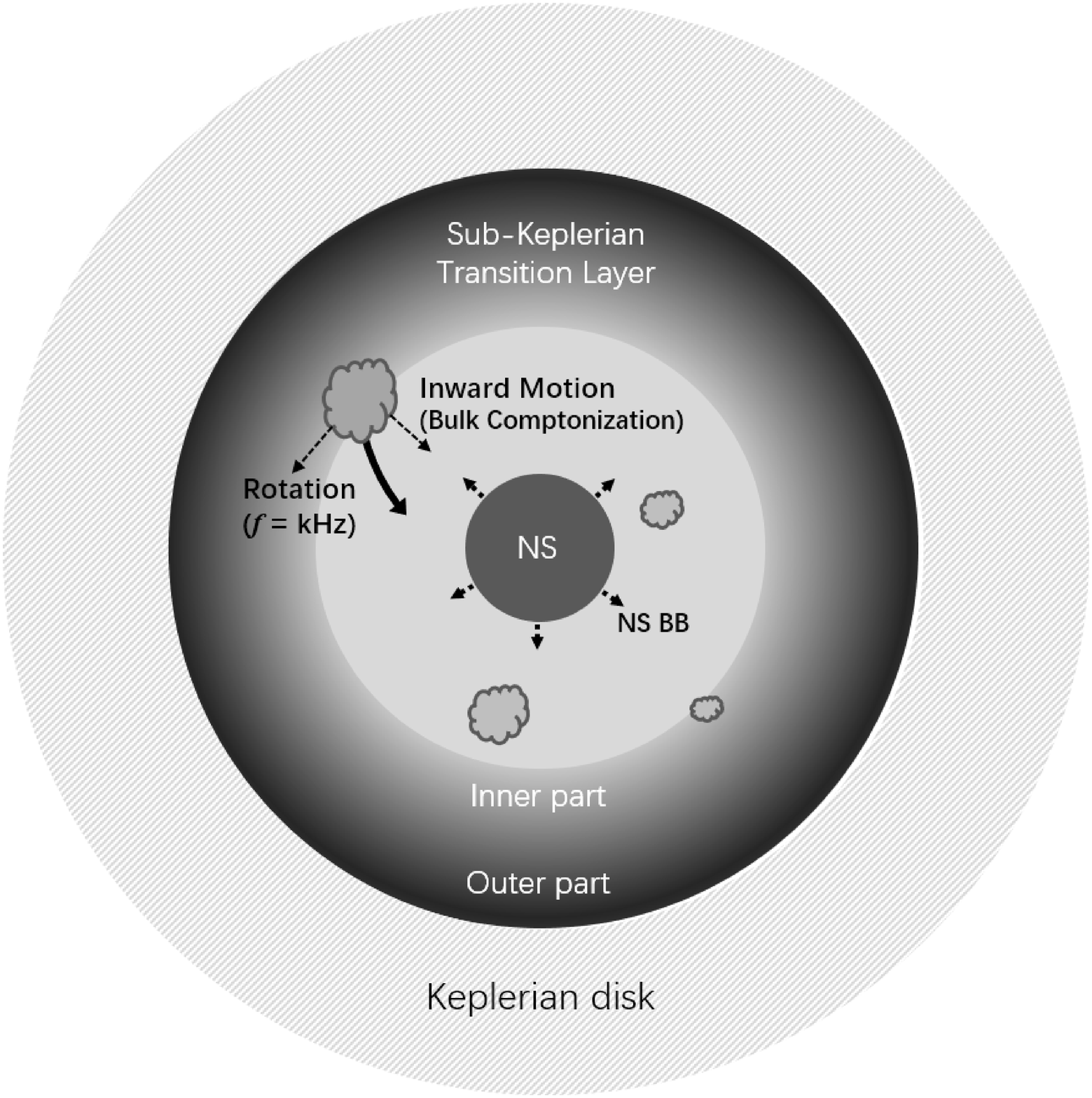}
  \end{minipage}%
  \caption{The schematic diagram of the origin of kHz QPOs in Sco X-1. The left panel is for the section view (adapted from \cite{Farinelli+2008}) and the right panel is for the top view. Clumps in the inner part of the transition layer rotate at sub-Keplerian frequency around the neutron star, which result in the kHz QPOs; the bulk Comptonization of the soft photons from the neutron star by the inward bulk motion of the clumps produces the high-energy photons.}
  \label{Fig:model}
\end{figure*}

\section{SUMMARY}

In this work, we presented a spectral-timing analysis for the kHz QPOs in Sco X-1 in a broad energy band with the data of $RXTE$ and $Insight$-HXMT. The upper and lower energy limits within which the emission has kHz QPOs are searched for the first time and the energy dependence of the kHz QPO properties on both the energy and the position in the $Z$--track are also investigated. Based on these results we discuss the origin of the kHz QPOs. The main results are summarized as follows:

1) kHz QPOs in Sco X-1 are detected on HB and NB by $RXTE$ and only detected on HB by $Insight$-HXMT.

2) The energy band with detectable kHz QPOs is determined for the first time with the method of sliding energy band: on HB it is $\sim$ 6.89--24.01 keV for the upper peaks and $\sim$ 8.68--21.78 keV for the lower peaks derived by $RXTE$, and $\sim$ 9--27.5 keV for the upper kHz QPOs by $Insight$-HXMT; on lower NB, the energy band is much narrower, $\sim$ 8.68--19.56 keV and $\sim$ 12.99--15.90 keV for the upper and lower peaks by $RXTE$, respectively.

3) Within the energy band, the centroid frequencies of kHz QPOs do not vary with energy, while the fractional rms increases steadily. Along the $Z$-track from HB to NB, the centroid frequencies of kHz QPOs in 8.68--12.99 keV increase from $\sim$ 900 Hz to $\sim$ 1100 Hz for the upper peaks and from $\sim$ 600 Hz to $\sim$ 900 Hz for the lower peaks, while the fractional rms decreases from 4\% to 2\% and from 3\% to 1\% respectively.

4) We derived the rms profile with the overlapping energy method to study the energy dependence of the fractional rms. It increases with energy and reaches a plateau at about 20 keV and 16 keV for the upper and lower peaks till $\sim$ 30 keV, though the error bars are large.
Our simulations show that $RXTE$ and $Insight$-HXMT are not sensitive enough at higher energy to detect the kHz QPO signal with the fraction rms of $\lesssim$ 10\%, but when the fractional rms increases to 14\% the signal can be detected by $Insight$-HXMT, implying that the fractional rms at higher energy may not increase with the same rate as in $\sim$ 6--20 keV. This also indicates that the upper energy limits of the kHz QPOs in Sco X-1 derived by $RXTE$ and $Insight$-HXMT are not so strict.

5) We analysed another set of $RXTE$ data covering much lower energy, and found that there is no detectable kHz QPO signal below $\sim$ 6 keV, indicating that the kHz QPOs could not come from the black body emission of the accretion disk and neutron star surface.

6) We fit the energy spectrum and the absolute rms spectrum with the two-Comptb model corresponding to photons from accretion disk and neutron star scattered off by the transition layer. Our results suggest that the kHz QPOs in Sco X-1 originate from the Comptonization of the inner part of the transition layer, where the rotation sets the frequency and the inward bulk motion makes the spectrum harder.

\acknowledgements {
This work is supported by the National Program on Key Research and Development Project (Grant No. 2016YFA0400801), the Bureau of International Cooperation, Chinese Academy of Sciences (GJHZ1864), the Strategic Pioneer Program on Space Science, Chinese Academy of Sciences (Grant No. XDA15310300) and the National Natural Science Foundation of China (Grant No. 11733009, 11673023, U2038104, U2031205, U1838201, U1838202, U1838111 and U1838115).
}

\end{document}